\newcommand{\bq}{\begin{equation}}
\newcommand{\eq}{\end{equation}}
\newcommand{\bqa}{\begin{eqnarray}}
\newcommand{\eqa}{\end{eqnarray}}
\newcommand{\fs}[1]{/\!\!\!#1}

\documentclass[a4paper]{article}
\usepackage{verbatim}
\usepackage{moreverb}
\usepackage{latexsym}
\usepackage{axodraw}
\usepackage{graphicx}
\usepackage{amsmath}

\title{Cancelling Quadratic Divergences Without Supersymmetry}
\author{M.T.M. van Kessel\footnote{Radboud Universiteit Nijmegen, Nijmegen, The Netherlands, M.vanKessel@science.ru.nl}}

\begin{document}

\maketitle

\section{Abstract}

We construct a theory which has the same particle content as softly broken minimal supersymmetric QED (MSQED) and is free of quadratic divergences up to two loops. Also this theory is completely gauge invariant. It appears that MSQED is not at all the only theory without these quadratic divergences. This proves that there exist non supersymmetric theories in which there are no quadratic divergences up to two loops.

\newpage

\section{Introduction}

In this paper we discuss an extension of QED where all quadratic divergences are absent up to two loop order. Of course the softly broken supersymmetric extension of QED (MSQED) hase this nice property. The Lagrangian of this theory can be constructed with the help of superfields (see e.g.\ ref. \cite{Wess}). In this way one can find the Feynman rules for MSQED. These Feynman rules are given in appendix 1. 

For supersymmetric theories one can prove (\cite{WessZum}, \cite{Capper}) that all Green's functions are free of quadratic divergences (QD). This is considered to be one of the great merits of supersymmetric theories, i.e.\ supersymmetric theories solve the naturalness problem. It is then natural to ask whether the specific supersymmetric theory one has constructed is the \underline{only} theory without QD. It is this question that we address in the case of MSQED.

In \cite{Veltman}, \cite{Inami}, \cite{Deshpande} and \cite{Jack} similar questions are addressed. Veltman \cite{Veltman} showed that in the Standard Model the QD vanish at 1-loop order if one imposes one constraint on the coupling constants.

Inami et al.\ \cite{Inami} investigate a model with one Majorana fermion and a number of real spin-0 fields. They demand that no QD occur up to 2-loop order. They find that for a theory with 2 or less spin-0 fields this condition uniquely determines supersymmetry, i.e.\ the supersymmetric theory is the only theory without QD. 

Deshpande et al.\ \cite{Deshpande} investigate several models and impose that all QD have to vanish at 1-loop order \underline{and} that the equations one gets for the coupling constants should be invariant under the renormalization group equation. They find, in all the cases they consider, that with these constraints one always gets the supersymmetric theory. Because they demand that their 1-loop equations are invariant under the renormalization group their calculations are actually of higher loop order, i.e.\ they demand that QD vanish at \underline{all} orders in perturbation theory.

Jack et al.\ \cite{Jack} investigate theories where, if the QD vanish at 1-loop order, the QD at 2-loop automatically vanish. They give an example of such a theory which is \underline{not} supersymmetric.

In \cite{Inami} and \cite{Deshpande} the absence of QD up to some loop order means the theory under consideration is necessarily supersymmetric. In this paper we show that this is not the case for all theories. Of course we do not claim that there exist non supersymmetric theories without QD at all orders in perturbation theory, i.e.\ we do \underline{not} claim that we can solve the naturalness problem without introducing supersymmetry. It is just interesting by itself how many orders one needs before one gets to a supersymmetric theory (if one gets there at all). As is shown in this article, in which we limit ourselves to 2-loop calculations, two loops is not enough to get to the supersymmetric theory. Also for this reason we \underline{don't} demand that our equations are invariant under the renormalization group. (as they do in e.g.\ \cite{Deshpande} and \cite{Kubo}.) If we would do so our calculation would effectively be of more than 2-loop order. To keep a clear sight on what happens at every order in perturbation theory we explicitly calculate amplitudes at every order. Of course practical reasons have disabled us to carry on with our analysis at 3-loop order and further

This paper is organized as follows. We start by postulating the particle content of our theory, which is the same as MSQED. Then we start constructing the vertices one by one. We will only introduce new vertices when it is absolutely necessary to keep everything free of QD. When introducing a new vertex we will keep the form general. Then we will fix it as much as possible by insisting that there are no QD. This process will be done up to 2-loop order.

After this construction up to 2-loop order we will see how much of the theory is actually fixed. It will appear that, to keep all Green's functions free of QD up to 2-loop order, the form of the vertices are fixed, however there is still a lot of freedom to choose the value of coupling constants. At this point we will also check that the theory we have built is gauge invariant.

\SetScale{0.5}

\section{Building the Theory}

We wish to build a theory with the same particle content as MSQED. This means we have one spin-${1\over2}$ fermion with charge $e$ and mass $m_e$ (electron $e$), one spin-1 boson with mass 0 (photon $\gamma$), two scalar particles with charge $e$ and mass $m_L$ and $m_R$ (left and right selectron $\tilde{e}_L$ and $\tilde{e}_R$) and one spin-${1\over2}$ Majorana fermion with charge 0 and mass $m_{\tilde{\gamma}}$ (photino $\tilde{\gamma}$). Because the photino is a Majorana fermion special Feynman rules have to be used. A convenient formulation of these Feynman rules is given by Denner et al. in \cite{Denner}. These rules are explained in appendix 1. Note that we work in the Feynman/Lorentz gauge. In section \ref{gaugeinv} we will come back to the point of gauge invariance of our theory.

Now we want to build a minimal theory with these particles which is gauge invariant, unitary and free of QD. Let's start by introducing the electron photon interaction we know from ordinary QED:
\bq
\begin{picture}(100, 30)(0, 3)
\ArrowLine(40, 10)(100, 10)
\ArrowLine(100, 10)(160,40)
\Photon(100, 10)(160, -20){-5}{3.5} 
\end{picture} \leftrightarrow \qquad -ie\gamma^{\mu}
\eq

\vspace{15pt}

If we only have this vertex we have QED with three extra particles which do not interact. We know then that the theory is gauge invariant and unitary, but \underline{not} free of QD. In the vacuum polarization at 1-loop order we encounter a QD. Only one diagram, $P_1^{\mu\nu}$, contributes to this vacuum polarization, and its QD part is easily calculated:
\bqa
P_1^{\mu\nu} &=&
\begin{picture}(60, 25)(0, 17)
\Photon(0, 40)(30, 40){5}{2.5}
\PText(5, 54)(0)[]{p}
\ArrowArcn(60, 40)(30, 180, 0)
\PText(60, 60)(0)[]{p-l}
\ArrowArcn(60, 40)(30, 0, 180)
\PText(60, 20)(0)[]{l}
\Photon(90, 40)(120, 40){5}{2.5}
\PText(115, 54)(0)[]{p}
\end{picture} \nonumber\\[15pt]
&=& -\frac{1}{(2 \pi)^4} \int d^4l \; \textrm{Tr}\left( (-ie) \gamma^{\nu} \frac{i(\fs{p}-\fs{l}+m_e)}{(p-l)^2 - m_e^2 + i \varepsilon} (-ie) \gamma^{\mu} \frac{i(-\fs{l}+m_e)}{l^2 - m_e^2 + i \varepsilon} \right) \nonumber\\
&\overset{\textrm{QD}}{\longrightarrow}& -{ie^2\over8\pi^2} \; \Lambda^2 \; g^{\mu\nu}
\eqa
In the last step we kept only the QD part. How we compute the QD part of a diagram is explained in appendix 2. We see that only the longitudinal part of the photon propagator contains a QD, the transversal part is free of QD because it is protected by the gauge invariance of the theory.

Here we see the need of another vertex with a photon. The only 4 possibilities are depicted below.\\[-20pt]
\begin{center}
\begin{picture}(100, 40)(0, 0)
\Line(40, 10)(100, 10)
\PText(70, 5)(0)[]{L}
\Line(100, 10)(160, 40)
\PText(135, 20)(0)[]{L}
\Photon(100, 10)(160, -20){-5}{3.5}
\end{picture}
\begin{picture}(100, 40)(0, 0)
\Line(40, 10)(100, 10)
\PText(70, 5)(0)[]{R}
\Line(100, 10)(160, 40)
\PText(135, 20)(0)[]{R}
\Photon(100, 10)(160, -20){-5}{3.5}
\end{picture}
\begin{picture}(100, 40)(0, 0)
\Line(40, 10)(100, 10)
\PText(70, 5)(0)[]{L}
\Line(100, 10)(160, 40)
\PText(135, 20)(0)[]{R}
\Photon(100, 10)(160, -20){-5}{3.5}
\end{picture}
\begin{picture}(100, 40)(0, 0)
\Line(40, 10)(100, 10)
\PText(70, 5)(0)[]{R}
\Line(100, 10)(160, 40)
\PText(135, 20)(0)[]{L}
\Photon(100, 10)(160, -20){-5}{3.5}
\end{picture}\\[20pt]
\end{center}
Now for the moment, to keep our theory as simple as possible let's only introduce the first two. The general form $E^{\mu}$ of these vertices is:
\bq \label{genexp}
E^{\mu} = Ap^{\mu}+Bq^{\mu} \;,
\eq
where $p^{\mu}$ is the momentum of the incoming selectron, $q^{\mu}$ is the momentum of the outgoing selectron and $A$ and $B$ are constants. All momenta are counted from left to right. Of course $A$ and $B$ could depend on $p^2$ and $q^2$, however to keep our theory as simple as possible we will not introduce form factors in the vertices. The exact form of the two vertices can be fixed with the Ward Takahashi identity (see e.g.\ \cite{Peskin}) for the 3-vertex at tree level:
\bq \label{WT}
\frac{i}{p^2 - m_L^2} E^{\mu} (q-p)_{\mu} \frac{i}{q^2 - m_L^2} = e\frac{i}{p^2 - m_L^2} - e\frac{i}{q^2 - m_L^2} \;.
\eq 
Substituting the general expression (\ref{genexp}) in (\ref{WT}) we find:
\bq
A = -ie, \quad B=-ie
\eq
so that both vertices become:\vspace*{-15pt}
\bq
\begin{picture}(100, 40)(0, 3)
\Line(40, 10)(100, 10)
\PText(70, 5)(0)[]{L}
\Line(100, 10)(160, 40)
\PText(135, 20)(0)[]{L}
\Photon(100, 10)(160, -20){-5}{3.5}
\end{picture} \leftrightarrow \qquad -ie(p+q)^{\mu} \quad, \quad
\begin{picture}(100, 40)(0, 3)
\Line(40, 10)(100, 10)
\PText(70, 5)(0)[]{R}
\Line(100, 10)(160, 40)
\PText(135, 20)(0)[]{R}
\Photon(100, 10)(160, -20){-5}{3.5}
\end{picture} \leftrightarrow \qquad -ie(p+q)^{\mu} \label{selphovert}
\eq

\vspace{15pt}

Now, to see if we still have unitarity consider the process $\tilde{e}_L \bar{\tilde{e}}_L \rightarrow \gamma \gamma$. The Ward identity (also see e.g.\ \cite{Peskin}) states that if we take one of the photons longitudinally polarized and off shell, but the other particles on shell, the matrix element should be zero. There are two diagrams:
\bq
M_1 = \quad
\begin{picture}(60, 40)(0, 18)
\Line(0, 0)(60, 20)
\PText(30, 2)(0)[]{L}
\PText(30, 13)(0)[]{_}
\PText(10, 20)(0)[]{p}
\PText(15, 15)(0)[]{2}
\Line(0, 80)(60, 60)
\PText(30, 65)(0)[]{L}
\PText(10, 95)(0)[]{p}
\PText(15, 90)(0)[]{1}
\Line(60, 20)(60, 60)
\PText(55, 40)(0)[]{L}
\Photon(60, 20)(120, 0){-5}{2.5}
\PText(110, 20)(0)[]{q}
\PText(115, 15)(0)[]{2}
\Photon(60, 60)(120, 80){5}{2.5}
\PText(110, 95)(0)[]{q}
\PText(115, 90)(0)[]{1}
\end{picture} \quad, \quad M_2 = \quad
\begin{picture}(60, 40)(0, 18)
\Line(0, 0)(60, 20)
\PText(30, 2)(0)[]{L}
\PText(30, 13)(0)[]{_}
\PText(10, 20)(0)[]{p}
\PText(15, 15)(0)[]{2}
\Line(0, 80)(60, 60)
\PText(30, 65)(0)[]{L}
\PText(10, 95)(0)[]{p}
\PText(15, 90)(0)[]{1}
\Line(60, 20)(60, 60)
\PText(55, 40)(0)[]{L}
\Photon(60, 20)(120, 80){-5}{3.5}
\PText(110, 20)(0)[]{q}
\PText(115, 15)(0)[]{2}
\Photon(60, 60)(120, 0){5}{3.5}
\PText(110, 95)(0)[]{q}
\PText(115, 90)(0)[]{1}
\end{picture}
\eq\\[15pt]
All indicated momenta are again flowing from left to right. Taking photon 1 longitudinally polarized ($\varepsilon_1 \simeq q_1$) and off shell a simple calculation gives:
\bq \label{M1plM2}
M_1 + M_2 = -2ie^2 \; q_1 \cdot \varepsilon_2 \;.
\eq
This means we need another diagram to satisfy the Ward identity. Now, it can easily be seen that none of the possible 3-vertices one can construct is going to help here. The simplest thing we can do then is introduce a new 4-vertex. It is clear that the only 4-vertex that will help is:
\begin{center}
\begin{picture}(100, 65)(0, 0)
\Line(40, 0)(100, 60)
\PText(70, 17)(0)[]{L}
\PText(70, 28)(0)[]{_}
\Line(40, 120)(100, 60)
\PText(70, 84)(0)[]{L}
\Photon(100, 60)(160, 0){-5}{4.5} 
\Photon(100, 60)(160, 120){5}{4.5} 
\end{picture}
\end{center}
This new vertex gives one extra diagram and it is easy to see from (\ref{M1plM2}) that it has to be of the form $A \; g^{\mu\nu}$, with $A=2ie^2$. Of course the same reasoning can be given for the process $\tilde{e}_R \bar{\tilde{e}}_R \rightarrow \gamma \gamma$ and we get two new 4-vertices:
\bq
\begin{picture}(100, 30)(0, 30)
\Line(40, 0)(100, 60)
\PText(70, 17)(0)[]{L}
\PText(70, 28)(0)[]{_}
\Line(40, 120)(100, 60)
\PText(70, 84)(0)[]{L}
\Photon(100, 60)(160, 0){-5}{4.5} 
\Photon(100, 60)(160, 120){5}{4.5} 
\end{picture} \leftrightarrow \qquad 2ie^2 \; g^{\mu\nu} \quad, \quad
\begin{picture}(100, 30)(0, 30)
\Line(40, 0)(100, 60)
\PText(70, 17)(0)[]{R}
\PText(70, 28)(0)[]{_}
\Line(40, 120)(100, 60)
\PText(70, 84)(0)[]{R}
\Photon(100, 60)(160, 0){-5}{4.5} 
\Photon(100, 60)(160, 120){5}{4.5} 
\end{picture} \leftrightarrow \qquad 2ie^2 \; g^{\mu\nu}
\eq\\[5pt]

Now we go back to the photon propagator. Because of the new vertices we get 4 new diagrams, of which the QD parts are readily calculated with the method outlined in appendix 2:
\bqa
P_2^{\mu\nu} &=& \quad
\begin{picture}(60, 25)(0, 17)
\Photon(0, 40)(30, 40){5}{2.5}
\PText(5, 54)(0)[]{p}
\CArc(60, 40)(30, 0, 180)
\PText(60, 80)(0)[]{L}
\PText(60, 60)(0)[]{p-l}
\CArc(60, 40)(30, 180, 0)
\PText(60, 3)(0)[]{L}
\PText(60, 20)(0)[]{l}
\Photon(90, 40)(120, 40){5}{2.5}
\PText(115, 54)(0)[]{p}
\end{picture} \quad \overset{\textrm{QD}}{\longrightarrow} -{ie^2\over16\pi^2} \; \Lambda^2 \; g^{\mu\nu} \nonumber\\
P_3^{\mu\nu} &=& \quad
\begin{picture}(60, 50)(0, 17)
\Photon(0, 40)(30, 40){5}{2.5}
\PText(5, 54)(0)[]{p}
\CArc(60, 40)(30, 0, 180)
\PText(60, 80)(0)[]{R}
\PText(60, 60)(0)[]{p-l}
\CArc(60, 40)(30, 180, 0)
\PText(60, 3)(0)[]{R}
\PText(60, 20)(0)[]{l}
\Photon(90, 40)(120, 40){5}{2.5}
\PText(115, 54)(0)[]{p}
\end{picture} \quad \overset{\textrm{QD}}{\longrightarrow} -{ie^2\over16\pi^2} \; \Lambda^2 \; g^{\mu\nu} \nonumber\\
P_4^{\mu\nu} &=& \quad
\begin{picture}(60, 50)(0, 17)
\Photon(0, 15)(120, 15){3}{6.5}
\PText(5, 28)(0)[]{p}
\PText(115, 28)(0)[]{p}
\CArc(60, 42)(25, -90, 270)
\PText(60, 75)(0)[]{L}
\PText(60, 55)(0)[]{l}
\end{picture} \quad \overset{\textrm{QD}}{\longrightarrow} {ie^2\over8\pi^2} \; \Lambda^2 \; g^{\mu\nu} \nonumber\\ 
P_5^{\mu\nu} &=& \quad
\begin{picture}(60, 50)(0, 17)
\Photon(0, 15)(120, 15){3}{6.5}
\PText(5, 28)(0)[]{p}
\PText(115, 28)(0)[]{p}
\CArc(60, 42)(25, -90, 270)
\PText(60, 75)(0)[]{R}
\PText(60, 55)(0)[]{l}
\end{picture} \quad \overset{\textrm{QD}}{\longrightarrow} {ie^2\over8\pi^2} \; \Lambda^2 \; g^{\mu\nu}
\eqa\\[5pt]
Note again that only the longitudinal parts of all diagrams get QD. We see now that all QD cancel in the sum $P_1^{\mu\nu}+\ldots+P_5^{\mu\nu}$. So with these 4 new vertices we solved the problems of the longitudinal part of the photon propagator.

Now consider the left selectron propagator. With the vertices we have introduced up to now we have two contributing diagrams:
\bqa
P_1 &=& \quad
\begin{picture}(60, 25)(0, 17)
\Line(0, 40)(30, 40)
\PText(15, 34)(0)[]{L}
\PText(5, 50)(0)[]{p}
\CArc(60, 40)(30, 0, 180)
\PText(60, 78)(0)[]{L}
\PText(60, 60)(0)[]{p-l}
\PhotonArc(60, 40)(30, 180, 0){-3}{5.5}
\PText(60, 20)(0)[]{l}
\Line(90, 40)(120, 40)
\PText(105, 34)(0)[]{L}
\PText(115, 50)(0)[]{p}
\end{picture} \quad \overset{\textrm{QD}}{\longrightarrow} \frac{ie^2}{16 \pi^2} \Lambda^2 \nonumber\\
P_2 &=& \quad
\begin{picture}(60, 50)(0, 17)
\Line(0, 15)(120, 15)
\PText(30, 8)(0)[]{L}
\PText(5, 25)(0)[]{p}
\PText(90, 8)(0)[]{L}
\PText(115, 25)(0)[]{p}
\PhotonArc(60, 40)(25, -90, 270){-5}{8.5}
\PText(60, 55)(0)[]{l}
\end{picture} \quad \overset{\textrm{QD}}{\longrightarrow} -\frac{ie^2}{4 \pi^2} \Lambda^2
\eqa\\
The QD do not cancel, so we need a new vertex again. It is easy to see that the only 3-vertices which will help here are vertices where a left selectron, an electron and a photino meet. There are four of these:\\[-25pt]
\begin{center}
\begin{picture}(100, 40)(0, 0)
\ArrowLine(40, 10)(100, 10)
\Line(100, 10)(160, 40)
\PText(135, 20)(0)[]{L}
\Line(100, 10)(160, -20)
\Photon(100, 10)(160, -20){-5}{3.5} 
\end{picture} 
\begin{picture}(100, 40)(0, 0)
\ArrowLine(40, 10)(100, 10)
\Line(100, 10)(160, 40)
\PText(135, 20)(0)[]{R}
\Line(100, 10)(160, -20)
\Photon(100, 10)(160, -20){-5}{3.5} 
\end{picture}
\begin{picture}(100, 40)(0, 0)
\Photon(40, 10)(100, 10){5}{3.5}
\Line(40, 10)(100, 10)
\Line(100, 10)(160, 40)
\PText(135, 18)(0)[]{L}
\PText(135, 29)(0)[]{_}
\ArrowLine(100, 10)(160, -20)
\end{picture} 
\begin{picture}(100, 40)(0, 0)
\Photon(40, 10)(100, 10){5}{3.5}
\Line(40, 10)(100, 10)
\Line(100, 10)(160, 40)
\PText(135, 18)(0)[]{R}
\PText(135, 29)(0)[]{_}
\ArrowLine(100, 10)(160, -20)
\end{picture}\\[20pt]
\end{center}
Because the Lagrangian should be Hermitian there are two relations between these four vertices:
\bqa
\begin{picture}(100, 25)(0, 3)
\Photon(40, 10)(100, 10){5}{3.5}
\Line(40, 10)(100, 10)
\Line(100, 10)(160, 40)
\PText(135, 18)(0)[]{L}
\PText(135, 29)(0)[]{_}
\ArrowLine(100, 10)(160, -20)
\end{picture} &=&
-\gamma^0 \left(\begin{picture}(100, 25)(0, 3)
\ArrowLine(40, 10)(100, 10)
\Line(100, 10)(160, 40)
\PText(135, 20)(0)[]{L}
\Line(100, 10)(160, -20)
\Photon(100, 10)(160, -20){-5}{3.5} 
\end{picture}\right)^{\dagger} \gamma^0 \nonumber\\
\begin{picture}(100, 25)(0, 3)
\Photon(40, 10)(100, 10){5}{3.5}
\Line(40, 10)(100, 10)
\Line(100, 10)(160, 40)
\PText(135, 18)(0)[]{R}
\PText(135, 29)(0)[]{_}
\ArrowLine(100, 10)(160, -20)
\end{picture} &=&
-\gamma^0 \left(\begin{picture}(100, 25)(0, 3)
\ArrowLine(40, 10)(100, 10)
\Line(100, 10)(160, 40)
\PText(135, 20)(0)[]{R}
\Line(100, 10)(160, -20)
\Photon(100, 10)(160, -20){-5}{3.5} 
\end{picture}\right)^{\dagger} \gamma^0
\eqa
Now because of these new vertices we get one new diagram:
\bq
P_3 = \quad
\begin{picture}(60, 25)(0, 17)
\Line(0, 40)(30, 40)
\PText(15, 34)(0)[]{L}
\PText(5, 50)(0)[]{p}
\ArrowArcn(60, 40)(30, 180, 0)
\PText(60, 60)(0)[]{p-l}
\PhotonArc(60, 40)(30, 180, 0){-3}{5.5}
\CArc(60, 40)(30, 180, 0)
\PText(60, 20)(0)[]{l}
\Line(90, 40)(120, 40)
\PText(105, 34)(0)[]{L}
\PText(115, 50)(0)[]{p}
\end{picture}
\eq\\[5pt]
Let's call the first of our new vertices $B_L(p,q)$:\\[-20pt]
\bq
B_L(p,q) = \quad
\begin{picture}(100, 40)(0, 3)
\ArrowLine(40, 10)(100, 10)
\Line(100, 10)(160, 40)
\PText(135, 20)(0)[]{L}
\Line(100, 10)(160, -20)
\Photon(100, 10)(160, -20){-5}{3.5} 
\end{picture}
\eq\\
where $p$ is the momentum in the electron line, flowing from left to right and $q$ is the momentum in the photino line, also flowing from left to right. Now the general expression for this vertex is a Fierz decomposition:
\bq \label{genexpB}
B_L(p,q) = S \mathbf{I} + V_{\alpha} \gamma^{\alpha} + T_{\alpha \beta} \sigma^{\alpha \beta} + A_{\alpha} \gamma^5 \gamma^{\alpha} + P \gamma^5
\eq
where $\sigma^{\alpha \beta}$ is defined as
\bq
\sigma^{\alpha \beta} = \frac{i}{2} ( \gamma^{\alpha} \gamma^{\beta} - \gamma^{\beta} \gamma^{\alpha} ) \;.
\eq
Since we do not want any form factors in our fundamental vertices, to keep the theory as simple as possible, $S$ and $P$ should be constants and for $V_{\alpha}$, $A_{\alpha}$ and $T_{\alpha \beta}$ we can write down the general expressions
\bqa
V_{\alpha} &=& v_1 \; p_{\alpha} + v_2 \; q_{\alpha} \nonumber\\
A_{\alpha} &=& a_1 \; p_{\alpha} + a_2 \; q_{\alpha} \nonumber\\
T_{\alpha \beta} &=& t_1 \; p_{\alpha} q_{\beta} + t_2 \; p_{\beta} q_{\alpha} + t_3 \; \epsilon_{\alpha \beta \delta \gamma} p^{\delta} q^{\gamma}
\eqa
where also $v_1$, $v_2$, $a_1$, $a_2$, $t_1$, $t_2$ and $t_3$ are constants. Notice we haven't included a term proportional to $g_{\alpha \beta}$ in $T_{\alpha \beta}$ since
\bq
g_{\alpha \beta} \; \sigma^{\alpha \beta} = 0 \;.
\eq
Now $P_3$ is given by
\bqa
P_3 &=& -\frac{1}{(2 \pi)^4} \int d^4l \; \textrm{Tr}\Bigg( B_L(p-l, -l) \frac{i(\fs{p}-\fs{l}+m_e)}{(p-l)^2 - m_e^2 + i \varepsilon} \cdot \nonumber\\
& & \phantom{-\frac{1}{(2 \pi)^4} \int d^4l \; \textrm{Tr}\Bigg(} \quad (-1) \gamma^0 B_L^{\dag}(p-l, -l) \gamma^0 \frac{i(-\fs{l}+m_{\tilde{\gamma}})}{l^2 - m_{\tilde{\gamma}}^2 + i \varepsilon} \Bigg) \label{P3}
\eqa
Note that we don't have to worry about the special Feynman rules for Majorana fermions here, since there is only one electron line the fermion flow can be fixed along the electron arrow unambiguously. Now the general expression (\ref{genexpB}) should be substituted in (\ref{P3}). This involves a lot of algebra and can best be done by computer. It appears we find even quartic divergences (i.e.\ proportional to $\Lambda^4$). The coefficients in front of these terms have to vanish, from which we find two equations:
\bqa
-3|t_1-t_2|^2-12|t_3|^2 &=& 0 \Rightarrow t_1=t_2, \quad t_3=0 \nonumber\\
4|v_1-v_2|^2+4|a_1-a_2|^2 &=& 0 \Rightarrow v_1=v_2, \quad a_1=a_2 \label{taveq}
\eqa
When we substitute these results in the coefficients for the QD parts proportional to $p^2$ and $(p^2)^2$ we find one more equation (the coefficient for $(p^2)^2 \Lambda^2$ is automatically zero when we substitute (\ref{taveq})):
\bq
|v_1|^2+|a_1|^2=0
\eq
which means that $v_1=v_2=a_1=a_2=0$. Note that because $t_1=t_2$ the whole third term in (\ref{genexpB}) vanishes because of the anti symmetry of $\sigma^{\alpha\beta}$.

So our general expression (\ref{genexpB}) is now reduced to
\bq
B_L = S \mathbf{I} + P \gamma^5 \;.
\eq
For $S$ and $P$ we get one constraint because the quadratically divergent part with no factor $(p^2)^n$ in front should cancel the QD in $P_1$ and $P_2$:
\bq \label{SPrel}
-3e^2 + 4(|S|^2 + |P|^2) = 0 \;.
\eq
 
Of course one can also consider the right selectron propagator and impose the absence of QD. Then a relation similar to (\ref{SPrel}) is found. So finally our 4 new vertices become:\\[-20pt]
\bqa
\begin{picture}(100, 40)(0, 3)
\ArrowLine(40, 10)(100, 10)
\Line(100, 10)(160, 40)
\PText(135, 20)(0)[]{L}
\Line(100, 10)(160, -20)
\Photon(100, 10)(160, -20){-5}{3.5} 
\end{picture} \quad &\leftrightarrow& S_L \mathbf{I} + P_L \gamma^5 \nonumber\\
\begin{picture}(100, 40)(0, 3)
\Photon(40, 10)(100, 10){5}{3.5}
\Line(40, 10)(100, 10)
\Line(100, 10)(160, 40)
\PText(135, 18)(0)[]{L}
\PText(135, 29)(0)[]{_}
\ArrowLine(100, 10)(160, -20)
\end{picture} \quad &\leftrightarrow& -S_L \mathbf{I} + P_L \gamma^5 \nonumber\\
\begin{picture}(100, 40)(0, 3)
\ArrowLine(40, 10)(100, 10)
\Line(100, 10)(160, 40)
\PText(135, 20)(0)[]{R}
\Line(100, 10)(160, -20)
\Photon(100, 10)(160, -20){-5}{3.5} 
\end{picture} \quad &\leftrightarrow& S_R \mathbf{I} + P_R \gamma^5 \nonumber\\
\begin{picture}(100, 40)(0, 3)
\Photon(40, 10)(100, 10){5}{3.5}
\Line(40, 10)(100, 10)
\Line(100, 10)(160, 40)
\PText(135, 18)(0)[]{R}
\PText(135, 29)(0)[]{_}
\ArrowLine(100, 10)(160, -20)
\end{picture} \quad &\leftrightarrow& -S_R \mathbf{I} + P_R \gamma^5
\eqa\\
with the constraints
\bqa
-3e^2 + 4(|S_L|^2 + |P_L|^2) &=& 0 \nonumber\\
-3e^2 + 4(|S_R|^2 + |P_R|^2) &=& 0 \;.
\eqa

To get one more constraint from one loop processes we can consider $\tilde{e}_L \rightarrow \tilde{e}_R$. There is only one diagram for this process:
\bq
P_1 = \quad
\begin{picture}(60, 25)(0, 18)
\Line(0, 40)(30, 40)
\PText(15, 34)(0)[]{L}
\PText(5, 50)(0)[]{p}
\ArrowArcn(60, 40)(30, 180, 0)
\PText(60, 60)(0)[]{p-l}
\PhotonArc(60, 40)(30, 180, 0){-3}{5.5}
\CArc(60, 40)(30, 180, 0)
\PText(60, 20)(0)[]{l}
\Line(90, 40)(120, 40)
\PText(105, 34)(0)[]{R}
\PText(115, 50)(0)[]{p}
\end{picture} \quad \overset{\textrm{QD}}{\longrightarrow} \quad (S_L^* S_R + P_L^* P_R) {i\over4\pi^2} \; \Lambda^2 \;,
\eq\\
so that we also get
\bq
S_L^* S_R + P_L^* P_R = 0 \;.
\eq

With the vertices we have up to now one can also calculate the QD in the electron and photino propagator. It appears that these have no QD, because of their fermionic character. Also diagrams with 3 or more external legs, which we haven't considered up to now, will have no QD, which can be verified by simple power counting.

So now we have constructed a theory free of QD up to 1-loop order. We still have freedom to choose our coupling constants $S_L$, $P_L$, $S_R$ and $P_R$ (8 real parameters against 4 real equations) and we haven't even seen the need for introducing the selectron 4-vertices that occur in MSQED. 

\section{QD at 2-Loop}

Now let's consider our theory at 2-loop order and impose the absence of QD. First consider the left selectron propagator. There are 29 1PI diagrams:
\bqa
& & P_1 = \quad
\begin{picture}(60, 25)(0, 17)
\Line(0, 40)(30, 40)
\Line(90, 40)(120, 40)
\PhotonArc(60, 40)(30, 180, 0){3}{6.5}
\CArc(60,40)(30, 0, 180)
\PhotonArc(60, 70)(15, 194, -14){3}{3.5}
\end{picture}
\qquad P_2 = \quad
\begin{picture}(60, 25)(0, 17)
\Line(0, 40)(30, 40)
\Line(90, 40)(120, 40)
\PhotonArc(60, 40)(30, 180, 0){3}{6.5}
\ArrowArcn(60,40)(30, 180, 0)
\PhotonArc(60, 70)(15, 194, -14){3}{2.5}
\CArc(60, 70)(15, 194, -14)
\end{picture}
\qquad P_3 = \quad
\begin{picture}(60, 35)(0, 17)
\Line(0, 40)(30, 40)
\Line(90, 40)(120, 40)
\PhotonArc(60, 40)(30, 180, 0){3}{6.5}
\CArc(60,40)(30, 0, 180)
\PhotonArc(60, 83)(13, -90, 270){2}{7.5}
\end{picture} \nonumber\\
& & P_4 = \quad
\begin{picture}(60, 50)(0, 17)
\Line(0, 40)(30, 40)
\Line(90, 40)(120, 40)
\CArc(60, 40)(30, 0, 180)
\Oval(60, 35)(12, 15)(0)
\Photon(30, 40)(45, 35){3}{2}
\Photon(75, 35)(90, 40){3}{2}
\PText(60, 18)(0)[]{L}
\end{picture}
\qquad P_5 = \quad
\begin{picture}(60, 50)(0, 17)
\Line(0, 40)(30, 40)
\Line(90, 40)(120, 40)
\CArc(60, 40)(30, 0, 180)
\Oval(60, 35)(12, 15)(0)
\Photon(30, 40)(45, 35){3}{2}
\Photon(75, 35)(90, 40){3}{2}
\PText(60, 18)(0)[]{R}
\end{picture}
\qquad P_6 = \quad
\begin{picture}(60, 50)(0, 17)
\Line(0, 40)(30, 40)
\Line(90, 40)(120, 40)
\CArc(60, 40)(30, 0, 180)
\Oval(60, 35)(12, 15)(0)
\ArrowLine(59, 47)(61, 47)
\ArrowLine(61, 23)(59, 23)
\Photon(30, 40)(45, 35){3}{2}
\Photon(75, 35)(90, 40){3}{2}
\end{picture} \nonumber\\
& & P_7 = \quad
\begin{picture}(60, 50)(0, 17)
\Line(0, 40)(30, 40)
\Line(90, 40)(120, 40)
\CArc(60, 40)(30, 0, 180)
\PhotonArc(60, 40)(30, 180, 0){-3}{5.5}
\Oval(60,-5)(12, 12)(0)
\PText(43, -5)(0)[]{L}
\end{picture}
\qquad P_8 = \quad
\begin{picture}(60, 50)(0, 17)
\Line(0, 40)(30, 40)
\Line(90, 40)(120, 40)
\CArc(60, 40)(30, 0, 180)
\PhotonArc(60, 40)(30, 180, 0){-3}{5.5}
\Oval(60,-5)(12, 12)(0)
\PText(43, -5)(0)[]{R}
\end{picture}
\qquad P_9 = \quad
\begin{picture}(60, 50)(0, 17)
\Line(0, 40)(30, 40)
\Line(90, 40)(120, 40)
\PhotonArc(60, 40)(30, 180, 0){3}{6.5}
\CArc(60, 40)(30, 180, 0)
\ArrowArcn(60,40)(30, 180, 0)
\ArrowLine(33, 52)(34, 54)
\ArrowLine(86, 54)(87, 52)
\PhotonArc(60, 70)(15, 194, -14){3}{3.5}
\end{picture} \nonumber\\
& & P_{10} = \quad
\begin{picture}(60, 50)(0, 17)
\Line(0, 40)(30, 40)
\Line(90, 40)(120, 40)
\PhotonArc(60, 40)(30, 180, 0){3}{6.5}
\CArc(60, 40)(30, 180, 0)
\CArc(60,40)(30, 0, 180)
\ArrowLine(33, 52)(34, 54)
\ArrowLine(86, 54)(87, 52)
\PhotonArc(60, 70)(15, 194, -14){3}{3.5}
\CArc(60, 70)(15, 194, -14)
\PText(60, 77)(0)[]{L}
\end{picture}
\qquad P_{11} = \quad
\begin{picture}(60, 50)(0, 17)
\Line(0, 40)(30, 40)
\Line(90, 40)(120, 40)
\PhotonArc(60, 40)(30, 180, 0){3}{6.5}
\CArc(60, 40)(30, 180, 0)
\CArc(60,40)(30, 0, 180)
\ArrowLine(33, 52)(34, 54)
\ArrowLine(86, 54)(87, 52)
\PhotonArc(60, 70)(15, 194, -14){3}{3.5}
\CArc(60, 70)(15, 194, -14)
\PText(60, 77)(0)[]{R}
\end{picture} 
\qquad P_{12} = \quad
\begin{picture}(60, 50)(0, 17)
\Line(0, 40)(30, 40)
\Line(90, 40)(120, 40)
\ArrowArcn(60,40)(30, 180, 0)
\Oval(60, 30)(15, 15)(0)
\ArrowLine(61, 45)(59, 45)
\Photon(30, 40)(45, 30){3}{1.5}
\Line(30, 40)(45, 30)
\Photon(75, 30)(90, 40){3}{1.5}
\Line(75, 30)(90, 40)
\PText(60, 10)(0)[]{L}
\end{picture} \nonumber\\
& & P_{13} = \quad
\begin{picture}(60, 50)(0, 17)
\Line(0, 40)(30, 40)
\Line(90, 40)(120, 40)
\ArrowArcn(60,40)(30, 180, 0)
\Oval(60, 30)(15, 15)(0)
\ArrowLine(61, 45)(59, 45)
\Photon(30, 40)(45, 30){3}{1.5}
\Line(30, 40)(45, 30)
\Photon(75, 30)(90, 40){3}{1.5}
\Line(75, 30)(90, 40)
\PText(60, 10)(0)[]{R}
\end{picture}
\qquad P_{14} = \quad
\begin{picture}(60, 50)(0, 17)
\Line(0, 40)(30, 40)
\Line(90, 40)(120, 40)
\ArrowArcn(60,40)(30, 180, 0)
\Oval(60, 30)(15, 15)(0)
\ArrowLine(59, 45)(61, 45)
\Photon(30, 40)(45, 30){3}{1.5}
\Line(30, 40)(45, 30)
\Photon(75, 30)(90, 40){3}{1.5}
\Line(75, 30)(90, 40)
\PText(60, 10)(0)[]{L}
\end{picture}
\qquad P_{15} = \quad
\begin{picture}(60, 50)(0, 17)
\Line(0, 40)(30, 40)
\Line(90, 40)(120, 40)
\ArrowArcn(60,40)(30, 180, 0)
\Oval(60, 30)(15, 15)(0)
\ArrowLine(59, 45)(61, 45)
\Photon(30, 40)(45, 30){3}{1.5}
\Line(30, 40)(45, 30)
\Photon(75, 30)(90, 40){3}{1.5}
\Line(75, 30)(90, 40)
\PText(60, 10)(0)[]{R}
\end{picture} \nonumber\\
& & P_{16} = \quad
\begin{picture}(60, 50)(0, 12)
\Line(0, 0)(120, 0)
\PhotonArc(60, 20)(20, -90, 47){-3}{4.5}
\PhotonArc(60, 20)(20, 133, 270){-3}{4.5}
\Oval(60, 40)(15, 15)(0)
\PText(60, 63)(0)[]{L}
\end{picture}
\qquad P_{17} = \quad
\begin{picture}(60, 50)(0, 12)
\Line(0, 0)(120, 0)
\PhotonArc(60, 20)(20, -90, 47){-3}{4.5}
\PhotonArc(60, 20)(20, 133, 270){-3}{4.5}
\Oval(60, 40)(15, 15)(0)
\PText(60, 63)(0)[]{R}
\end{picture} 
\qquad P_{18} = \quad
\begin{picture}(60, 50)(0, 12)
\Line(0, 0)(120, 0)
\PhotonArc(60, 20)(20, -90, 47){-3}{4.5}
\PhotonArc(60, 20)(20, 133, 270){-3}{4.5}
\Oval(60, 40)(15, 15)(0)
\ArrowLine(59, 55)(61, 55)
\ArrowLine(61, 25)(59, 25)
\end{picture} \nonumber\\
& & P_{19} = \quad
\begin{picture}(60, 50)(0, 12)
\Line(0, 0)(120, 0)
\PhotonArc(60, 15)(15, -90, 270){-3}{9.5}
\Oval(60, 48)(15, 15)(0)
\PText(80, 48)(0)[]{L}
\end{picture}
\qquad P_{20} = \quad
\begin{picture}(60, 50)(0, 12)
\Line(0, 0)(120, 0)
\PhotonArc(60, 15)(15, -90, 270){-3}{9.5}
\Oval(60, 48)(15, 15)(0)
\PText(80, 48)(0)[]{R}
\end{picture} \nonumber\\
& & P_{21} = \quad
\begin{picture}(60, 50)(0, 17)
\Line(0, 40)(30, 40)
\Line(90, 40)(120, 40)
\Line(60, 10)(60, 70)
\CArc(60, 40)(30, 180, 270)
\CArc(60, 40)(30, 0, 90)
\PhotonArc(60, 40)(30, 90, 180){3}{3.5}
\PhotonArc(60, 40)(30, 270, 360){3}{3.5}
\end{picture}
\qquad P_{22} = \quad
\begin{picture}(60, 50)(0, 17)
\Line(0, 40)(30, 40)
\Line(90, 40)(120, 40)
\Line(60, 10)(60, 70)
\CArc(60, 40)(30, 180, 270)
\CArc(60, 40)(30, 0, 90)
\PhotonArc(60, 40)(30, 90, 180){3}{3.5}
\PhotonArc(60, 40)(30, 270, 360){3}{3.5}
\CArc(60, 40)(30, 270, 360)
\ArrowLine(60, 39)(60, 41)
\ArrowLine(80, 62)(82, 60)
\end{picture}
\qquad P_{23} = \quad
\begin{picture}(60, 50)(0, 17)
\Line(0, 40)(30, 40)
\Line(90, 40)(120, 40)
\Line(60, 10)(60, 70)
\CArc(60, 40)(30, 180, 270)
\CArc(60, 40)(30, 0, 90)
\PhotonArc(60, 40)(30, 90, 180){3}{3.5}
\CArc(60, 40)(30, 90, 180)
\PhotonArc(60, 40)(30, 270, 360){3}{3.5}
\ArrowLine(60, 39)(60, 41)
\ArrowLine(38, 20)(40, 18)
\end{picture} \nonumber\\
& & P_{24} = \quad
\begin{picture}(60, 50)(0, 17)
\Line(0, 40)(30, 40)
\Line(90, 40)(120, 40)
\Line(60, 10)(60, 70)
\PText(55, 40)(0)[]{L}
\CArc(60, 40)(30, 180, 270)
\CArc(60, 40)(30, 0, 90)
\PhotonArc(60, 40)(30, 90, 180){3}{3.5}
\CArc(60, 40)(30, 90, 180)
\PhotonArc(60, 40)(30, 270, 360){3}{3.5}
\CArc(60, 40)(30, 270, 360)
\ArrowLine(38, 20)(40, 18)
\ArrowLine(80, 62)(82, 60)
\end{picture}
\qquad P_{25} = \quad
\begin{picture}(60, 50)(0, 17)
\Line(0, 40)(30, 40)
\Line(90, 40)(120, 40)
\Line(60, 10)(60, 70)
\PText(55, 40)(0)[]{R}
\CArc(60, 40)(30, 180, 270)
\CArc(60, 40)(30, 0, 90)
\PhotonArc(60, 40)(30, 90, 180){3}{3.5}
\CArc(60, 40)(30, 90, 180)
\PhotonArc(60, 40)(30, 270, 360){3}{3.5}
\CArc(60, 40)(30, 270, 360)
\ArrowLine(38, 20)(40, 18)
\ArrowLine(80, 62)(82, 60)
\end{picture}
\qquad P_{26} = \quad
\begin{picture}(60, 50)(0, 17)
\Line(0, 40)(120, 40)
\PhotonArc(60, 40)(30, 180, 360){3}{6.5}
\PhotonArc(45, 40)(15, 0, 180){3}{3.5}
\end{picture} \nonumber\\
& & P_{27} = \quad
\begin{picture}(60, 50)(0, 17)
\Line(0, 40)(120, 40)
\PhotonArc(60, 40)(30, 180, 360){3}{6.5}
\PhotonArc(75, 40)(15, 0, 180){3}{3.5}
\end{picture}
\qquad P_{28} = \quad
\begin{picture}(60, 50)(0, 17)
\Line(0, 40)(20, 40)
\Line(100, 40)(120, 40)
\CArc(40, 40)(20, 0, 180)
\CArc(80, 40)(20, 0, 180)
\PhotonArc(40, 40)(20, 180, 360){3}{4.5}
\PhotonArc(80, 40)(20, 180, 360){3}{4.5}
\end{picture}
\qquad P_{29} = \quad
\begin{picture}(60, 50)(0, 17)
\Line(0, 40)(30, 40)
\Line(90, 40)(120, 40)
\CArc(60, 40)(30, 0, 180)
\Photon(30, 40)(90, 40){3}{5.5}
\PhotonArc(60, 40)(30, 180, 360){3}{6.5}
\end{picture}
\eqa\\[15pt]
Note that for these 2-loops diagrams the fact that the photino is a Majorana fermion is important. Otherwise diagrams 14, 15, 24 and 25 wouldn't have been possible at all. It is also for these diagrams that the special Feynman rules of the Majorana fermion become important because the arrows of the electron lines clash.

Because of the large number of diagrams we have automatized the computation of the QD in the propagators. The program \verb+LOOPS.frm+, written in \verb+FORM+ \cite{Vermaseren} can do the calculation of the QD for us. This code generates all diagrams at two loop order (also the 1PR ones) and keeps only the QD parts. The QD parts are written in terms of the standard integrals $A$ and $B$:
\bqa
A &=& {1\over(2\pi)^4} \int d^4l_1 \; d^4l_2 \; \frac{1}{l_1^2+i \varepsilon} \; \frac{1}{(l_1-l_2)^2+i \varepsilon} \;\frac{1}{l_2^2+i \varepsilon} \nonumber\\
B &=& {1\over(2\pi)^4} \int d^4l_1 \; d^4l_2 \; \frac{1}{(l_1^2+i \varepsilon)^2} \; \frac{1}{l_2^2+i \varepsilon}
\eqa
Notice that we omitted the masses, since these are unimportant for the QD part of the integral, integrals $A$ (or $B$) with different masses in the propagators can just be added as far as their QD part is concerned. Also the external momentum $p$ can be put to zero in the denominators, since they are also unimportant for the QD parts.

Running the program in the case of the selectron propagator gives us the coefficients in front of $A$ and $B$. It appears that the coefficient of $B$ already vanishes if we substitute the constraints we found at one loop order. The $A$-part gives a new constraint, its coefficient is:
\bqa
0 &=& 11ie^4 - 2ie^2 |S_L|^2 - 2ie^2 |P_L|^2 - 10i |S_L|^4 - 10i |P_L|^4 - 16i |S_L|^2 |P_L|^2 + \nonumber\\
& & 2i (S_L^*)^2 P_L^2 + 2i S_L^2 (P_L^*)^2 - 10i |S_L|^2 |S_R|^2 - 10i |P_L|^2 |P_R|^2 + \nonumber\\
& & -2i |S_L|^2 |P_R|^2 - 2i |P_L|^2 |S_R|^2 + 2i S_L^* P_L S_R^* P_R + 2i S_L P_L^* S_R P_R^* + \nonumber\\
& & -6i S_L^* P_L S_R P_R^* - 6i S_L P_L^* S_R^* P_R \label{2looplsel}
\eqa

We could now proceed with the other propagators at 2-loop order, however, first there is a problem. Consider the three constraints we found at one loop order again:
\bqa
3e^2 - 4|S_L|^2 - 4|P_L|^2 &=& 0 \nonumber\\
3e^2 - 4|S_R|^2 - 4|P_R|^2 &=& 0 \nonumber\\
S_L S_R^* + P_L P_R^* &=& 0 \;. \label{1loopbef}
\eqa
From the second equation we see that
\bq
|S_R|^2 = \frac{3}{4} e^2 - |P_R|^2 \;.
\eq
From the last one it follows that
\bq \label{PL}
P_L = -\frac{S_L S_R^*}{P_R^*} \;.
\eq
Substituting both relations in the first equation in (\ref{1loopbef}) gives:
\bq
|S_L| = |P_R| \;.
\eq
From (\ref{PL}) it then follows that
\bq
|S_R| = |P_L| \;.
\eq
If we use all these results in the 2-loop condition (\ref{2looplsel}) this condition reduces to:
\bq
\frac{11}{4} ie^4 = 0 \;.
\eq
Apparently our constraints are incompatible. This forces us to introduce new vertices. The only possible 3-vertices we can still introduce are:\\[-20pt]
\begin{center}
\begin{picture}(100, 40)(0, 0)
\Line(40, 10)(100, 10)
\PText(70, 5)(0)[]{L}
\Line(100, 10)(160, 40)
\PText(135, 20)(0)[]{R}
\Photon(100, 10)(160, -20){-5}{3.5}
\end{picture} \qquad
\begin{picture}(100, 40)(0, 0)
\Line(40, 10)(100, 10)
\PText(70, 5)(0)[]{R}
\Line(100, 10)(160, 40)
\PText(135, 20)(0)[]{L}
\Photon(100, 10)(160, -20){-5}{3.5}
\end{picture}\\[20pt]
\end{center}
With the Ward Takahashi identities these vertices can be fixed like we fixed (\ref{selphovert}), they will have exactly the same expression as (\ref{selphovert}). But this means we will ruin the photon propagator at one loop order again!

So we have to introduce new 4-vertices. Since the problem occurs with the selectrons it seems plausible to introduce selectron 4-vertices. These are also the simplest 4-vertices we can introduce because of the scalar character of the selectrons. So we will introduce:
\bq
\begin{picture}(100, 40)(0, 30)
\Line(40, 0)(160, 120)
\PText(70, 20)(0)[]{L}
\PText(130, 84)(0)[]{L}
\Line(40, 120)(160, 0)
\PText(70, 84)(0)[]{L}
\PText(130, 20)(0)[]{L}
\end{picture} \quad = -iV_L \qquad
\begin{picture}(100, 40)(0, 30)
\Line(40, 0)(160, 120)
\PText(70, 20)(0)[]{R}
\PText(130, 84)(0)[]{R}
\Line(40, 120)(160, 0)
\PText(70, 84)(0)[]{R}
\PText(130, 20)(0)[]{R}
\end{picture} \quad = -iV_R
\eq\\[15pt]
\bq
\begin{picture}(100, 40)(0, 30)
\Line(40, 0)(160, 120)
\PText(70, 20)(0)[]{R}
\PText(130, 84)(0)[]{L}
\Line(40, 120)(160, 0)
\PText(70, 84)(0)[]{L}
\PText(130, 20)(0)[]{R}
\end{picture} \quad = -iV_{LR}
\eq\\[15pt]

Because of these new vertices the left and right selectron propagator at 1-loop order also change. It is easy to see that our new 1-loop constraints become:
\bqa
3e^2 - 4|S_L|^2 - 4|P_L|^2 + V_L + V_{LR} &=& 0 \nonumber\\
3e^2 - 4|S_R|^2 - 4|P_R|^2 + V_R + V_{LR} &=& 0 \nonumber\\
S_L S_R^* + P_L P_R^* &=& 0 \;. \label{1loop}
\eqa
Nothing changes in the electron, photon and photino propagator at one loop order. 

The new constraint for the selectron propagator at two loop order can also calculated with \verb+LOOPS.frm+ again. We get:
\bqa
0 &=& 11ie^4 - 2ie^2 |S_L|^2 - 2ie^2 |P_L|^2 - 10i |S_L|^4 - 10i |P_L|^4 - 16i |S_L|^2 |P_L|^2 + \nonumber\\
& & 2i (S_L^*)^2 P_L^2 + 2i S_L^2 (P_L^*)^2 - 10i |S_L|^2 |S_R|^2 - 10i |P_L|^2 |P_R|^2 + \nonumber\\
& & -2i |S_L|^2 |P_R|^2 - 2i |P_L|^2 |S_R|^2 + 2i S_L^* P_L S_R^* P_R + 2i S_L P_L^* S_R P_R^* + \nonumber\\
& & -6i S_L^* P_L S_R P_R^* - 6i S_L P_L^* S_R^* P_R - 2ie^2 V_L - 2ie^2 V_{LR} + \nonumber\\
& & 2i |S_L|^2 V_L + 2i |P_L|^2 V_L + 2i |S_R|^2 V_{LR} + 2i |P_R|^2 V_{LR} + \nonumber\\
& & \frac{1}{2} i V_L^2 + i V_{LR}^2 \label{2loop1}
\eqa
The QD part of the right selectron propagator and the process $\tilde{e}_L \rightarrow \tilde{e}_R$ at two loop order can also be calculated. From these we get:
\bqa
0 &=& 11ie^4 - 2ie^2 |S_R|^2 - 2ie^2 |P_R|^2 - 10i |S_R|^4 - 10i |P_R|^4 - 16i |S_R|^2 |P_R|^2 + \nonumber\\
& & 2i (S_R^*)^2 P_R^2 + 2i S_R^2 (P_R^*)^2 - 10i |S_R|^2 |S_L|^2 - 10i |P_R|^2 |P_L|^2 + \nonumber\\
& & -2i |S_R|^2 |P_L|^2 - 2i |P_R|^2 |S_L|^2 + 2i S_R^* P_R S_L^* P_L + 2i S_R P_R^* S_L P_L^* + \nonumber\\
& & -6i S_R^* P_R S_L P_L^* - 6i S_R P_R^* S_L^* P_L - 2ie^2 V_R - 2ie^2 V_{LR} + \nonumber\\
& & 2i |S_R|^2 V_R + 2i |P_R|^2 V_R + 2i |S_L|^2 V_{LR} + 2i |P_L|^2 V_{LR} + \nonumber\\
& & \frac{1}{2} i V_R^2 + i V_{LR}^2 \label{2loop2}
\eqa
\bqa
0 &=& -2ie^2 S_L^* S_R - 2ie^2 P_L^* P_R - 8i |S_L|^2 P_L^* P_R - 10i |S_L|^2 S_L^* S_R + \nonumber\\
& & 2i S_L (P_L^*)^2 S_R - 8i |P_L|^2 S_L^* S_R - 8i |P_R|^2 S_L^* S_R -10i |S_R|^2 S_L^* S_R + \nonumber\\
& & 2i S_L^* S_R^* P_R^2 + 2i (S_L^*)^2 P_L P_R - 10i |P_L|^2 P_L^* P_R - 8i |S_R|^2 P_L^* P_R + \nonumber\\
& & 2i P_L^* S_R^2 P_R^* - 10i |P_R|^2 P_L^* P_R + 2i S_L^* S_R V_{LR} + 2i P_L^* P_R V_{LR} \label{2loop3}
\eqa
Of course (\ref{2loop2}) is just (\ref{2loop1}) with $L$ and $R$ swapped.

Also the QD parts of the electron, photon and photino propagator can be calculated with \verb+LOOPS.frm+. These appear to be zero already, independent of our choice of coupling constants.

\section{1 Loop Conclusions}

Using our constraints from 1 loop (\ref{1loop}) we can eliminate 4 of our 11 free (real) parameters. Let's define:
\bqa
S_L &\equiv& |S_L| \; e^{i\phi_L} \nonumber\\
P_L &\equiv& |P_L| \; e^{i\psi_L} \nonumber\\
S_R &\equiv& |S_R| \; e^{i\phi_R} \nonumber\\
P_R &\equiv& |P_R| \; e^{i\psi_R} \;.
\eqa
Now we use (\ref{1loop}) to express everything in the 3 phases $\phi_R$, $\psi_L$ and $\psi_R$, the magnitude $|P_R|$ and the 4-vertex constants $V_L$, $V_R$ and $V_{LR}$.
\bqa
\phi_L &=& \phi_R+\psi_L-\psi_R+\pi+2\pi n \nonumber\\[5pt]
|S_L|^2 &=& \alpha |P_R|^2 \nonumber\\
|P_L|^2 &=& \alpha \left( {3\over4}e^2-|P_R|^2+{1\over4}V_R+{1\over4}V_{LR} \right) \nonumber\\
|S_R|^2 &=& {3\over4}e^2-|P_R|^2+{1\over4}V_R+{1\over4}V_{LR} \label{1looprels}
\eqa
Here $\alpha$ is defined as
\bq
\alpha \equiv \frac{{3\over4}e^2+{1\over4}V_L+{1\over4}V_{LR}}{{3\over4}e^2+{1\over4}V_R+{1\over4}V_{LR}}
\eq
and the inequalities
\bqa
V_L+V_{LR} &\geq& -3e^2 \nonumber\\
V_R+V_{LR} &\geq& -3e^2
\eqa
should be satisfied in order that there exists a solution at all.

We see there is still a lot of freedom left if one only wants a theory free of QD up to 1-loop order.

\section{2 Loop Conclusions}

Now we can use the 1 loop relations (\ref{1looprels}) in the 2 loop constraints (\ref{2loop1}), (\ref{2loop2}) and (\ref{2loop3}). The easiest way is to use the 1-loop relations in (\ref{2loop3}) first. Then this equation simplifies to
\bq
\left( S_L S_R - P_L P_R \right) \left( 1-{1\over\alpha} \right) \left( \left(P_L^*\right)^2 - \left(S_L^*\right)^2 \right) = 0 \;.
\eq
We can satisfy this equation in 3 ways:
\begin{enumerate}
\item $S_L S_R = P_L P_R$
\item $\alpha = 1$
\item $P_L^* = \pm S_L^*$
\end{enumerate}

\subsection{Case 1}

In this case (\ref{2loop1}) and (\ref{2loop2}) simplify to two quadratic equations in $V_L$, $V_R$ and $V_{LR}$, it appears that $|P_R|$ drops out. We can use these two equations to express $V_L$ and $V_R$ in terms of $V_{LR}$. This gives 4 solutions. We are left with 4 real degrees of freedom:
\bq
\psi_L, \quad \psi_R, \quad |P_R|, \quad V_{LR}
\eq
The 4 solutions are:
\bqa
\varphi_L &=& \psi_L + {1\over2}\pi + \pi(m+n) \nonumber\\
\varphi_R &=& \psi_R - {1\over2}\pi + \pi(m-n) \nonumber\\
|S_L|^2 &=& |P_R|^2 \nonumber\\
|P_L|^2 = |S_R|^2 &=& {7\over2}e^2 - |P_R|^2 + {1\over2}V_{LR} \pm {1\over4} \sqrt{110e^4+44e^2V_{LR}-2V_{LR}^2} \nonumber\\
V_L = V_R &=& 11e^2 + V_{LR} \pm \sqrt{110e^4+44e^2V_{LR}-2V_{LR}^2} \label{case11}
\eqa
and
\bqa
\varphi_L &=& \psi_L + {1\over2}\pi + \pi(m+n) \nonumber\\
\varphi_R &=& \psi_R - {1\over2}\pi + \pi(m-n) \nonumber\\
|S_L|^2 &=& \alpha |P_R|^2 \nonumber\\
|P_L|^2 &=& \alpha \left( {7\over3}e^2 - |P_R|^2 + {2\over3}V_{LR} \mp {1\over12} \sqrt{397e^4+226e^2V_{LR}-11V_{LR}^2} \right) \nonumber\\
|S_R|^2 &=& {7\over3}e^2 - |P_R|^2 + {2\over3}V_{LR} \mp {1\over12} \sqrt{397e^4+226e^2V_{LR}-11V_{LR}^2} \nonumber\\
V_L &=& {19\over3}e^2 + {5\over3}V_{LR} \pm {1\over3} \sqrt{397e^4+226e^2V_{LR}-11V_{LR}^2} \nonumber\\
V_R &=& {19\over3}e^2 + {5\over3}V_{LR} \mp {1\over3} \sqrt{397e^4+226e^2V_{LR}-11V_{LR}^2} \label{case12}
\eqa
with
\bq
\alpha = \frac{28e^2+8V_{LR}\pm\sqrt{397e^4+226e^2V_{LR}-11V_{LR}^2}}{28e^2+8V_{LR}\mp\sqrt{397e^4+226e^2V_{LR}-11V_{LR}^2}} \;.
\eq

\subsection{Case 2}

In this case (\ref{2loop1}) and (\ref{2loop2}) simplify to the same equation, where also $|P_R|$ drops out. This equation has two solutions in terms of $V_{LR}$. We are left with 5 real degrees of freedom:
\bq
\varphi_R, \quad \psi_L, \quad \psi_R, \quad |P_R|, \quad V_{LR}
\eq
The two solutions are:
\bqa
\varphi_L &=& \varphi_R + \psi_L - \psi_R + \pi + 2\pi n \nonumber\\
|S_L|^2 &=& |P_R|^2 \nonumber\\
|P_L|^2 = |S_R|^2 &=& {7\over2}e^2 - |P_R|^2 + {1\over2}V_{LR} \pm {1\over4} \sqrt{110e^4+44e^2V_{LR}-2V_{LR}^2} \nonumber\\
V_L = V_R &=& 11e^2 + V_{LR} \pm \sqrt{110e^4+44e^2V_{LR}-2V_{LR}^2} \label{case2}
\eqa
We see that the two solutions in this case are like the first two solutions of case 1 (\ref{case11}), just slightly more general (Here $\varphi_R$ is also free.).

\subsection{Case 3}

In this case we also get two quadratic equations in $V_L$, $V_R$ and $V_{LR}$, which also have four solutions. We are left with 3 real degrees of freedom:
\bq
\psi_L, \quad \psi_R, \quad V_{LR}
\eq
and the four solutions are:
\bqa
\varphi_L &=& \psi_L + \pi m \nonumber\\
\varphi_R &=& \psi_R - \pi + \pi m \nonumber\\
|S_L|^2 = |P_L|^2 = |S_R|^2 = |P_R|^2 &=& {7\over4}e^2 + {1\over4}V_{LR} \pm {1\over8} \sqrt{110e^4+44e^2V_{LR}-2V_{LR}^2} \nonumber\\
V_L = V_R &=& 11e^2 + V_{LR} \pm \sqrt{110e^4+44e^2V_{LR}-2V_{LR}^2} \label{case31}
\eqa
and
\bqa
\varphi_L &=& \psi_L + \pi m \nonumber\\
\varphi_R &=& \psi_R - \pi + \pi m \nonumber\\
|S_L|^2 = |P_L|^2 &=& {7\over8}e^2 + {1\over4}V_{LR} \pm {1\over24} \sqrt{183e^4+132e^2V_{LR}-6V_{LR}^2} \nonumber\\
|S_R|^2 = |P_R|^2 &=& {7\over8}e^2 + {1\over4}V_{LR} \mp {1\over24} \sqrt{183e^4+132e^2V_{LR}-6V_{LR}^2} \nonumber\\
V_L &=& 4e^2 + V_{LR} \pm {1\over3} \sqrt{183e^4+132e^2V_{LR}-6V_{LR}^2} \nonumber\\
V_R &=& 4e^2 + V_{LR} \mp {1\over3} \sqrt{183e^4+132e^2V_{LR}-6V_{LR}^2} \label{case32}
\eqa

\begin{figure}[h]
\begin{center}
\includegraphics[width=14cm]{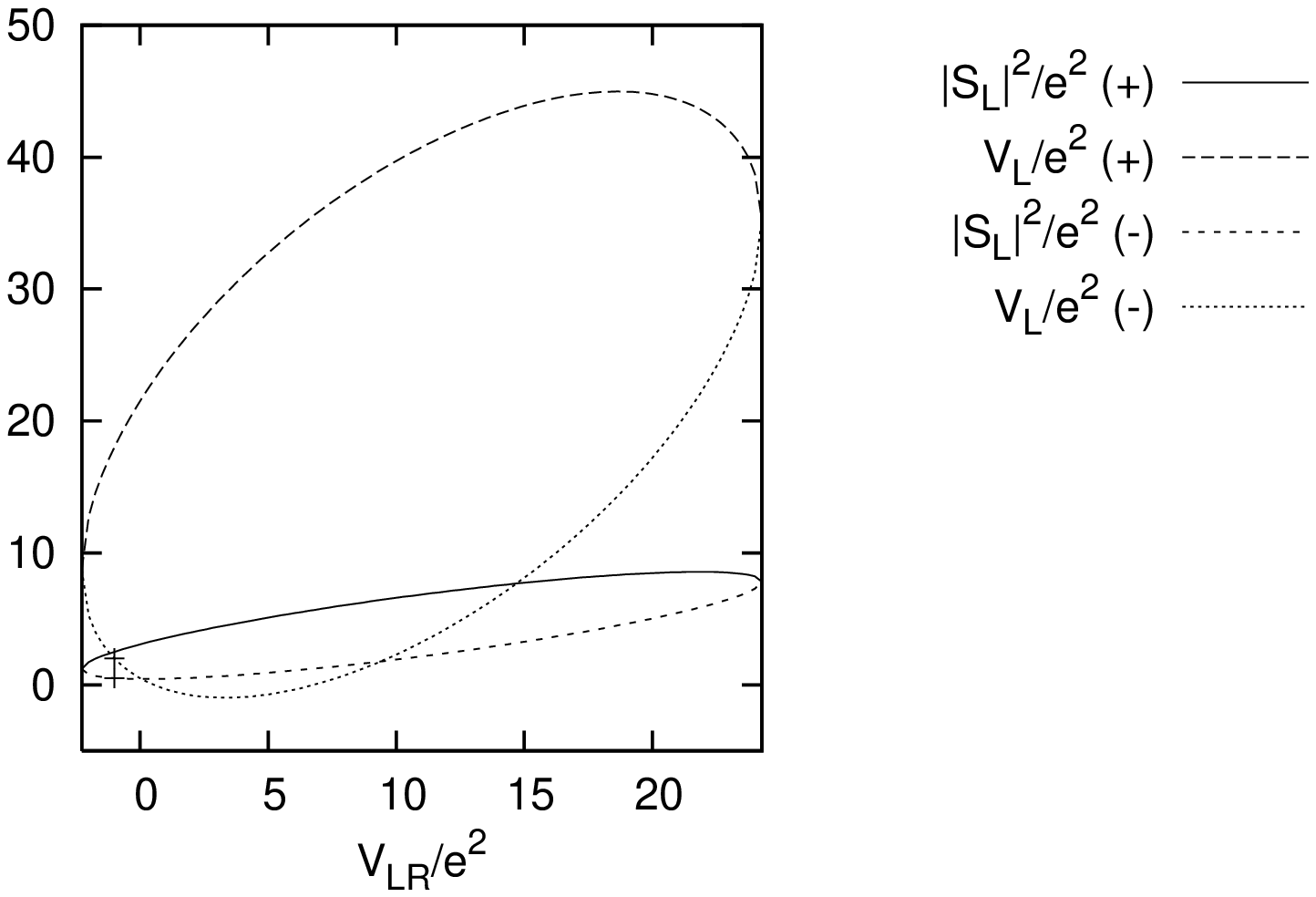}
\end{center}
\caption{The solutions (\ref{case31}). $|S_L|^2/e^2=|P_L|^2/e^2=|S_R|^2/e^2=|P_R|^2/e^2$ And $V_L/e^2=V_R/e^2$ are plotted as functions of $V_{LR}/e^2$. Both the solution with (+) and (-) are shown. The MSQED solution is indicated by the crosses.}
\label{plotcase31}
\end{figure}

The first two solutions are plotted in figure \ref{plotcase31}. In this figure the crosses indicate the values for the MSQED coupling constants, they lie on the curves of the (-) solution (\ref{case31}).

The final conclusion is that, in the case of MSQED, MSQED is not at all the only theory free of QD up to two loops. With the demand that up to two loops all QD vanish one can fix the form of the vertices in a minimal extension of QED. For the values of the coupling constants there is still a lot of freedom left.

\section{Gauge Invariance}\label{gaugeinv}

While building our theory we have fixed some of our vertices by using the Ward or Ward Takahashi identities. We chose our vertices such that the processes under consideration satisfied these identities. This does not mean however that our complete theory is gauge invariant. To see that our complete theory is gauge invariant for general $S_L$, $P_L$, $S_R$, $P_R$, $V_L$, $V_R$ and $V_{LR}$ we will demonstrate this at the level of the Lagrangian density.
From our Feynman rules the Lagrangian density can be read off. We find:
\bq
\mathcal{L} = \mathcal{L}_{K} + \mathcal{L}_{I} \;,
\eq
with
\bqa
\mathcal{L}_{K} &=& - \frac{1}{4} F^{\mu \nu} F_{\mu \nu} - \frac{1}{2} \bar{\psi}_{\tilde{\gamma}} \left(-i\gamma^{\mu} \partial_{\mu} + m_{\tilde{\gamma}}\right) \psi_{\tilde{\gamma}} - \tilde{e}^{\dag}_{L} (\Box + m_L^2) \tilde{e}_{L} - \tilde{e}^{\dag}_{R} (\Box + m_R^2) \tilde{e}_{R} + \nonumber\\
& & -\bar{\psi}_{e} \left(-i\gamma^{\mu} \partial_{\mu} + m_e\right) \psi_{e} \;, \nonumber\\[5pt]
\mathcal{L}_{I} &=& -e A^{\mu} \bar{\psi}_{e} \gamma_{\mu} \psi_{e} + \nonumber\\
& & -ie A^{\mu} \left(\tilde{e}^{\dag}_{L} \partial_{\mu} \tilde{e}_{L} - \left(\partial_{\mu} \tilde{e}^{\dag}_{L}\right) \tilde{e}_{L}\right) - ie A^{\mu} \left(\tilde{e}^{\dag}_{R} \partial_{\mu} \tilde{e}_{R} - \left(\partial_{\mu} \tilde{e}^{\dag}_{R}\right) \tilde{e}_{R}\right) + \nonumber\\
& & e^2 A^2 \tilde{e}^{\dag}_{L} \tilde{e}_{L} + e^2 A^2 \tilde{e}^{\dag}_{R} \tilde{e}_{R} + \nonumber\\
& & -i \; \tilde{e}_{L} \bar{\psi}_{e} \left(-S_L^*+P_L^*\gamma^5\right) \psi_{\tilde{\gamma}} - i \; \tilde{e}^{\dag}_{L} \bar{\psi}_{\tilde{\gamma}} \left(S_L+P_L\gamma^5\right) \psi_{e} + \nonumber\\
& & -i \; \tilde{e}_{R} \bar{\psi}_{e} \left(-S_R^*+P_R^*\gamma^5\right) \psi_{\tilde{\gamma}} - i \; \tilde{e}^{\dag}_{R} \bar{\psi}_{\tilde{\gamma}} \left(S_R+P_R\gamma^5\right) \psi_{e} + \nonumber\\
& & -\frac{1}{4} V_L \; \tilde{e}^{\dag}_{L} \tilde{e}_{L} \tilde{e}^{\dag}_{L} \tilde{e}_{L} - \frac{1}{4} V_R \; \tilde{e}^{\dag}_{R} \tilde{e}_{R} \tilde{e}^{\dag}_{R} \tilde{e}_{R} - V_{LR} \; \tilde{e}^{\dag}_{L} \tilde{e}_{L} \tilde{e}^{\dag}_{R} \tilde{e}_{R} \;,
\eqa
where of course
\bq
F^{\mu\nu} \equiv \partial^{\mu} A^{\nu} - \partial^{\nu} A^{\mu} \;.
\eq

Now it is easy to check that indeed this Lagrangian density is invariant under the gauge transformation:
\bqa
A^{\mu}(x) &\rightarrow& A^{\mu}(x) + \partial^{\mu} \Lambda(x) \nonumber\\
\psi_e(x) &\rightarrow& \exp\left(-ie\Lambda(x)\right) \psi_e(x) \nonumber\\
\tilde{e}_L(x) &\rightarrow& \exp\left(-ie\Lambda(x)\right) \tilde{e}_L \nonumber\\
\tilde{e}_R(x) &\rightarrow& \exp\left(-ie\Lambda(x)\right) \tilde{e}_R
\eqa
(Strictly speaking it is not this Lagrangian density that is invariant, but the action $S=\int d^4x \; \mathcal{L}$, the Lagrangian density itself is invariant up to terms which are total derivatives.)

\newpage

\appendix
\section{Appendix 1: The Feynman rules for MSQED}

In the three figures below the Feynman rules for MSQED are given in the Feynman/Lorentz gauge. The straight line with arrow indicates the electron, the combination of a straight line and a wiggly line indicates the photino, the straight lines without arrow with a label L or R indicate the left or right selectron and the wiggly line indicates the photon. Here the photino is a Majorana fermion, which makes it impossible to fix a fermion flow in each diagram unambiguously. To be able to work with Feynman rules we follow Denner et al. \cite{Denner}. They state that for any fermion loop one should choose an arbitrary orientation for the fermion flow. One should count the momenta along this chosen fermion flow and the vertices $\Gamma$ change as:
\bq
\Gamma \longrightarrow C\Gamma^T C^{-1}.
\eq
Here $T$ denotes transposition and $C$ is the charge-conjugation matrix. The expression on the right hand side is equal to
\bq
C\Gamma^T C^{-1} = \eta\Gamma
\eq
where $\eta$ is given by:
\bq
\eta = \left\{ \begin{array}{ll}
1 \quad & \textrm{for} \quad \Gamma = \mathbf{I}, i \gamma^5, \gamma^{\mu} \gamma^5 \\
-1 \quad & \textrm{for} \quad \Gamma = \gamma^{\mu}, \sigma^{\mu \nu}
\end{array} \right.
\eq 
The fermion flow in all diagrams will be indicated by an arrow next to the fermion line.

\SetScale{0.5}

\begin{figure}[!h]
\caption{The Feynman rules for the bare propagators in MSQED. The indicated momenta flow from left to right. The arrows next to the fermion lines represent the chosen fermion flow.}
\label{f1}
 
\hspace{100pt}
\begin{picture}(100, 30)(0, 4)
\ArrowLine(40, 10)(160, 10) 
\LongArrow(80, 0)(120, 0)
\PText(100, 30)(0)[]{p}
\end{picture} 
$\leftrightarrow \qquad \frac{i(\fs{p}+m_e)}{p^2-m_e^2+i \varepsilon}$

\hspace{100pt}
\begin{picture}(100, 30)(0, 4)
\ArrowLine(40, 10)(160, 10) 
\LongArrow(120, 0)(80, 0)
\PText(100, 30)(0)[]{p}
\end{picture} 
$\leftrightarrow \qquad \frac{i(-\fs{p}+m_e)}{p^2-m_e^2+i \varepsilon}$

\hspace{100pt}
\begin{picture}(100, 30)(0, 4)
\Photon(40, 10)(160, 10){5}{5.5}
\Line(40, 10)(160, 10) 
\LongArrow(80, 0)(120, 0)
\PText(100, 30)(0)[]{p}
\end{picture} 
$\leftrightarrow \qquad \frac{i(\fs{p}+m_{\tilde{\gamma}})}{p^2-m_{\tilde{\gamma}}^2+i \varepsilon}$

\hspace{100pt}
\begin{picture}(100, 30)(0, 4)
\Photon(40, 10)(160, 10){5}{5.5}
\Line(40, 10)(160, 10) 
\LongArrow(120, 0)(80, 0)
\PText(100, 30)(0)[]{p}
\end{picture} 
$\leftrightarrow \qquad \frac{i(-\fs{p}+m_{\tilde{\gamma}})}{p^2-m_{\tilde{\gamma}}^2+i \varepsilon}$

\hspace{100pt}
\begin{picture}(100, 30)(0, 4)
\Line(40, 10)(160, 10)
\PText(100, 0)(0)[]{L} 
\PText(100, 30)(0)[]{p}
\end{picture} 
$\leftrightarrow \qquad \frac{i}{p^2-m_L^2+i \varepsilon}$

\hspace{100pt}
\begin{picture}(100, 30)(0, 4)
\Line(40, 10)(160, 10)
\PText(100, 0)(0)[]{R} 
\PText(100, 30)(0)[]{p}
\end{picture} 
$\leftrightarrow \qquad \frac{i}{p^2-m_R^2+i \varepsilon}$

\hspace{100pt}
\begin{picture}(100, 30)(0, 4)
\Photon(40, 10)(160, 10){5}{5.5}
\PText(100, 30)(0)[]{p}
\end{picture} 
$\leftrightarrow \qquad \frac{-ig_{\mu \nu}}{p^2+i \varepsilon}$

\end{figure}

\begin{figure}[!h]
\caption{The Feynman rules for the 3-vertices in MSQED. The indicated momenta flow from left to right. The arrows next to the fermion lines represent the chosen fermion flow. Notice that the electron-selectron-photino vertices where the fermion flow is opposite to the electron arrow are not explicitly given, these vertices have the same expression as their counterparts where the fermion flow is along the electron arrow.}
\label{f2}

\hspace{100pt}
\begin{picture}(100, 40)(0, 4)
\ArrowLine(40, 10)(100, 10)
\ArrowLine(100, 10)(160,40)
\Photon(100, 10)(160, -20){-5}{3.5} 
\LongArrowArc(85, 60)(40, 270, 310)
\end{picture} 
$\leftrightarrow \qquad -ie \gamma^{\mu}$

\hspace{100pt}
\begin{picture}(100, 40)(0, 4)
\ArrowLine(40, 10)(100, 10)
\ArrowLine(100, 10)(160,40)
\Photon(100, 10)(160, -20){-5}{3.5} 
\LongArrowArcn(85, 60)(40, 310, 270)
\end{picture} 
$\leftrightarrow \qquad ie \gamma^{\mu}$

\hspace{100pt}
\begin{picture}(100, 40)(0, 4)
\Line(40, 10)(100, 10)
\Line(100, 10)(160, 40)
\PText(70, 5)(0)[]{L}
\PText(135, 20)(0)[]{L}
\PText(70, 30)(0)[]{p}
\PText(130, 45)(0)[]{q}
\Photon(100, 10)(160, -20){-5}{3.5} 
\end{picture} 
$\leftrightarrow \qquad -ie (p+q)^{\mu}$

\hspace{100pt}
\begin{picture}(100, 40)(0, 4)
\Line(40, 10)(100, 10)
\Line(100, 10)(160, 40)
\PText(70, 5)(0)[]{R}
\PText(135, 20)(0)[]{R}
\PText(70, 30)(0)[]{p}
\PText(130, 45)(0)[]{q}
\Photon(100, 10)(160, -20){-5}{3.5} 
\end{picture} 
$\leftrightarrow \qquad -ie (p+q)^{\mu}$

\hspace{100pt}
\begin{picture}(100, 40)(0, 4)
\ArrowLine(40, 10)(100, 10)
\Photon(100, 10)(160, 40){-5}{3.5}
\Line(100, 10)(160, 40)
\PText(135, 5)(0)[]{L}
\Line(100, 10)(160, -20)
\LongArrowArc(85, 60)(40, 270, 310)
\end{picture} 
$\leftrightarrow \qquad \frac{e}{\sqrt{2}} ( \mathbf{I} + \gamma^5 )$

\hspace{100pt}
\begin{picture}(100, 40)(0, 4)
\Line(40, 10)(100, 10)
\PText(70, 5)(0)[]{L}
\ArrowLine(100, 10)(160, 40)
\Line(100, 10)(160, -20)
\Photon(100, 10)(160, -20){-5}{3.5} 
\LongArrowArcn(160, 10)(12, 240, 120)
\end{picture} 
$\leftrightarrow \qquad -\frac{e}{\sqrt{2}} ( \mathbf{I} - \gamma^5 )$

\hspace{100pt}
\begin{picture}(100, 40)(0, 4)
\ArrowLine(40, 10)(100, 10)
\Photon(100, 10)(160, 40){-5}{3.5}
\Line(100, 10)(160, 40)
\PText(135, 5)(0)[]{R}
\Line(100, 10)(160, -20)
\LongArrowArc(85, 60)(40, 270, 310)
\end{picture} 
$\leftrightarrow \qquad \frac{e}{\sqrt{2}} ( \mathbf{I} - \gamma^5 )$

\hspace{100pt}
\begin{picture}(100, 40)(0, 4)
\Line(40, 10)(100, 10)
\PText(70, 5)(0)[]{R}
\ArrowLine(100, 10)(160, 40)
\Line(100, 10)(160, -20)
\Photon(100, 10)(160, -20){-5}{3.5} 
\LongArrowArcn(160, 10)(12, 240, 120)
\end{picture} 
$\leftrightarrow \qquad -\frac{e}{\sqrt{2}} ( \mathbf{I} + \gamma^5 )$

\end{figure}

\begin{figure}[!h]
\caption{The Feynman rules for the 4-vertices in MSQED.}
\label{f3}

\hspace{100pt}
\begin{picture}(100, 65)(0, 30)
\Line(40, 0)(100, 60)
\PText(70, 17)(0)[]{L}
\PText(70, 28)(0)[]{_}
\Line(40, 120)(100, 60)
\PText(70, 84)(0)[]{L}
\Photon(100, 60)(160, 0){-5}{4.5} 
\Photon(100, 60)(160, 120){5}{4.5} 
\end{picture} 
$\leftrightarrow \qquad 2ie^2 g^{\mu \nu}$

\hspace{100pt}
\begin{picture}(100, 65)(0, 30)
\Line(40, 0)(100, 60)
\PText(70, 17)(0)[]{R}
\PText(70, 28)(0)[]{_}
\Line(40, 120)(100, 60)
\PText(70, 84)(0)[]{R}
\Photon(100, 60)(160, 0){-5}{4.5} 
\Photon(100, 60)(160, 120){5}{4.5} 
\end{picture} 
$\leftrightarrow \qquad 2ie^2 g^{\mu \nu}$

\hspace{100pt}
\begin{picture}(100, 65)(0, 30)
\Line(40, 0)(160, 120)
\PText(70, 20)(0)[]{L}
\PText(130, 84)(0)[]{L}
\Line(40, 120)(160, 0)
\PText(70, 84)(0)[]{L}
\PText(130, 20)(0)[]{L}
\end{picture} 
$\leftrightarrow \qquad -2ie^2$

\hspace{100pt}
\begin{picture}(100, 65)(0, 30)
\Line(40, 0)(160, 120)
\PText(70, 20)(0)[]{R}
\PText(130, 84)(0)[]{R}
\Line(40, 120)(160, 0)
\PText(70, 84)(0)[]{R}
\PText(130, 20)(0)[]{R}
\end{picture} 
$\leftrightarrow \qquad -2ie^2$

\hspace{100pt}
\begin{picture}(100, 65)(0, 30)
\Line(40, 0)(160, 120)
\PText(70, 20)(0)[]{R}
\PText(130, 84)(0)[]{L}
\Line(40, 120)(160, 0)
\PText(70, 84)(0)[]{L}
\PText(130, 20)(0)[]{R}
\end{picture} 
$\leftrightarrow \qquad ie^2$

\end{figure}
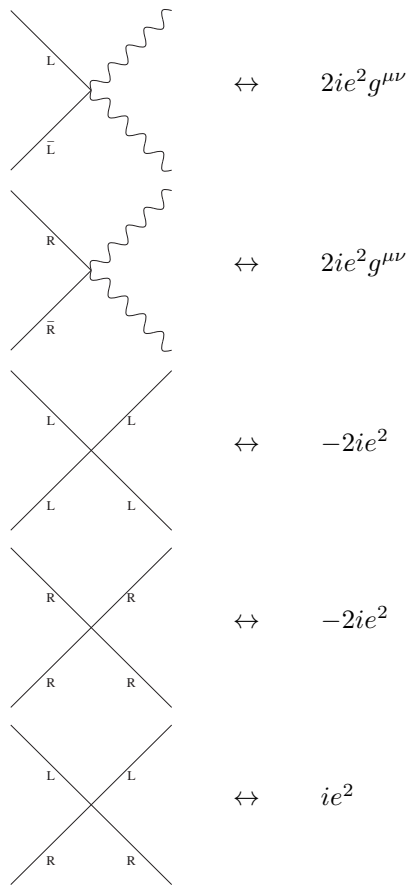
\clearpage

\section{Appendix 2: Computing QD parts}

To see how a loop integral diverges one has to introduce a regulator, in our case a cut-off momentum. This cut-off can be a strict cut-off at momentum $\Lambda$, which means that no momentum occurring in a bare propagator may exceed $\Lambda$. In our case it is however more convenient to work with a smoother cut-off momentum (\cite{KleppeWoodard}, \cite{EvensMoffat}) introduced by
\bq
\frac{1}{l_E^2 + m^2} \rightarrow \int_{\Lambda^{-2}}^\infty d\alpha \; e^{-\alpha(l_E^2+m^2)}
\eq
where $l_E$ is a Euclidean momentum. This means that in all our loop integrals we have to do a Wick rotation first, to get a Euclidean form.

The loop integrals (1 and 2 loop) for propagator diagrams which we have to know will in general depend on the momentum $p$ flowing through the propagator and several masses. However, if one is only interested in the worst divergent part of a loop integral these can in principal all be set to zero. We shall do this for the momentum $p$, but not for the masses, to avoid introducing infrared singularities. On can however take all masses equal.

Now consider the 1-loop integral
\bq
I_n = \int d^4l \; \frac{1}{(l^2 - m^2 + i \varepsilon)^n} \;.
\eq
After the Wick rotation one finds:
\bq
I_n = i(-1)^n \int d^4l_E \; \frac{1}{(l_E^2 + m^2)^n}
\eq
Using the regularization scheme introduced above this becomes:
\bq
I_n = i\pi^2(-1)^n \int_{\Lambda^{-2}}^{\infty} d\alpha_1 \ldots d\alpha_n \; \frac{1}{\left(\alpha_1+\ldots+\alpha_n\right)^2} \; e^{-\left(\alpha_1+\ldots+\alpha_n\right)m^2}
\eq
From this expression it is easy to see the worst divergent parts:
\bqa
I_1 &=& -i\pi^2 \; \Lambda^2 \nonumber\\
I_2 &=& 2i\pi^2 \; \ln \Lambda \nonumber\\
I_{n>2} &=& i\pi^2(-1)^n \frac{1}{(n-2)(n-1)} \frac{1}{m^{2n-4}} \;,
\eqa
indeed verifying what one expects from simple power counting.

The only 2-loop integral which we will encounter is:
\bq
I = \int d^4l_1 d^4l_2 \; \frac{1}{l_1^2-m^2+i \varepsilon} \; \frac{1}{(l_1-l_2)^2-m^2+i \varepsilon} \; \frac{1}{l_2^2-m^2+i \varepsilon} \;.
\eq
After Wick rotating, using the regularization and performing the momentum integrals this becomes
\bq
I = \pi^4 \int_{\Lambda^{-2}}^{\infty} d\alpha_1 d\alpha_2 d\alpha_3 \; \frac{1}{(\alpha_1 \alpha_2 + \alpha_1 \alpha_3 + \alpha_2 \alpha_3)^2} \; e^{-\left(\alpha_1+\alpha_2+\alpha_3\right)m^2} \;.
\eq
The worst divergent part (in this case quadratically divergent, as we will see below) can be obtained by putting the exponential to 1. Then our 2-loop integral becomes
\bq
I = \pi^4 \ln\left({64\over27}\right) \; \Lambda^2 \;,
\eq
indeed showing a QD.

\end{document}